# Distance-based Protein Folding Powered by Deep Learning

Jinbo Xu

*Toyota Technological Institute at Chicago*

*6045 S Kenwood, IL, 60637, USA*

*jinboxu@gmail.com*

Contact-assisted protein folding has made very good progress, but two challenges remain. One is accurate contact prediction for proteins lack of many sequence homologs and the other is that time-consuming folding simulation is often needed to predict good 3D models from predicted contacts. We show that protein distance matrix can be predicted well by deep learning and then directly used to construct 3D models without folding simulation at all. Using distance geometry to construct 3D models from our predicted distance matrices, we successfully folded 21 of the 37 CASP12 hard targets with a median family size of 58 effective sequence homologs within 4 hours on a Linux computer of 20 CPUs. In contrast, contacts predicted by direct coupling analysis (DCA) cannot fold any of them in the absence of folding simulation and the best CASP12 group folded 11 of them by integrating predicted contacts into complex, fragment-based folding simulation. The rigorous experimental validation on 15 CASP13 targets show that among the 3 hardest targets of new fold our distance-based folding servers successfully folded 2 large ones with <150 sequence homologs while the other servers failed on all three, and that our ab initio folding server also predicted the best, high-quality 3D model for a large homology modeling target. Further experimental validation in CAMEO shows that our ab initio folding server predicted correct fold for a membrane protein of new fold with 200 residues and 229 sequence homologs while all the other servers failed. These results imply that deep learning offers an efficient and accurate solution for ab initio folding on a personal computer.

## Introduction

De novo protein structure prediction from sequence alone is one of the most challenging problems in computational biology and the progress on this problem has been slow for a while. Nevertheless, in recent years good progress has been achieved thanks to accurate contact prediction enabled by direct coupling analysis (DCA)[1-9] and deep convolutional neural networks (DCNN)[10-16]. As such, contact-assisted protein folding has gained a lot of attention and lots of effort have been devoted to contact prediction.

We have developed the CASP12-winning method RaptorX-Contact[10] that uses deep convolutional residual neural network (ResNet) to predict contacts. ResNet is one type of DCNN[17], but much more powerful than traditional or plain DCNN. RaptorX-Contact has good accuracy even for some proteins with only dozens of effective sequence homologs. The accuracy of RaptorX-Contact decreases much more slowly than DCA when more predicted contacts are evaluated even if protein under study has thousands of sequence homologs (see Table 1 in the RaptorX-Contact paper[10]). As reported in[10, 12], without folding simulation, the 3D models constructed from contacts predicted by RaptorX-Contact have much better accuracy than those built from contacts predicted by DCA methods such as CCMpred[6] and the CASP11 winner MetaPSICOV[20] (equivalent to a neural network of 3 hidden layers).

Both ResNet and DCA are global prediction methods because they predict whether one pair of residues form a contact or not by considering the status of the other residue pairs, which is the key to the significant improvement in contact prediction. However, ResNet is better than DCA in that 1) ResNet may capture higher-order residue correlation while DCA mainly focuses on pairwise relationship; 2) ResNet tries to learn the global context of a protein contact matrix and uses it to predict the status of one residue pair; and 3) existing DCA methods are roughly a linear model with tens of millions of parameters to be estimated from a single protein family while ResNet is a nonlinear model with fewer parameters to be estimated from thousands of protein families. Before ResNet, deep learning (DL) such as CMAPpro[18] and Deep Belief Networks (DBN)[19] were used for contact prediction, but ResNet is the first DL method that greatly outperforms DCA and shallow methods such as MetaPSICOV[20]. Different from ResNet and DCA, both DBN and MetaPSICOV are local prediction methods because when predicting the relationship of one residue pair they ignore the status of the other pairs. Cheng group, who proposed DBN, has switched to DCNN for contact prediction[13]. Traditional DCNN was used by Peng group in CASP12[16], but he switched to ResNet in CASP13. Inspired by the success of ResNet in CASP12, most CASP13 contact prediction groups have applied ResNet or DCNN to contact prediction[13, 15, 21], as shown in the CASP13 abstract book.

While contact prediction is drawing lots of attention, here we study distance prediction and treat contact prediction as a by-product. Distance matrix contains not only finer-grained information than contact matrix, but also more physical constraints of a protein structure, e.g., distance is metric while contact is not. A distance matrix can determine a protein structure (except mirror image) much more accurately than a contact matrix. Trained by distance instead of contact matrices, ResNet may automatically learn more about the intrinsic properties of a protein structure and thus, greatly reduce conformation space and improve folding accuracy. Further, different from DCA that aims to predict only a small number of contacts and then use them to assist folding simulation, we would like to predict the whole distance matrix and then to directly construct protein 3D models without folding simulation at all, which may significantly reduce running time needed for protein folding, especially for a large protein. As we use many more distance restraints to construct 3D models, the impact of distance prediction error may be reduced (by the law of large numbers). In contrast, contact-assisted folding simulation may be misguided by several wrongly-predicted contacts and needs a long time to generate a good conformation for a large protein.

Although there are very few studies on distance prediction, it is not totally new. For example, Aszodi et al[22] predicted inter-residue distance from pairwise conserved hydrophobicity score and used it to reconstruct 3D models of a protein. Kloczkowski et al[23] described a spectral decomposition method for inter-residue distance prediction. We employed a probabilistic neural network to predict inter-residue distance and then derived position-specific distance potential from predicted distance[24]. Pietal et al. [25] predicted inter-residue distance matrix from a contact matrix and then built protein 3D models from predicted distance. Kukic et al.[26] proposed a recursive neural network method for inter-residue distance prediction and studied distance-based protein folding, but the predicted distance has a large error and the resultant 3D models have poor quality. Recently, we showed that predicted inter-residue distance may improve protein threading by a large margin[27] when only weakly similar templates are available.

In addition to inter-atom distance prediction, we also employ deep ResNet to predict secondary structure and backbone torsion angles, although they are much less important than distance for folding. By feeding these three types of predicted restraints to CNS[28], a computer program for experimental protein structure determination, without folding simulation we can construct more accurate 3D models than what can be achieved from predicted contacts, as evidenced by our experiments on a set of 37 CASP12 hard targets[14] and a set of 41 CAMEO hard targets[10]. Our distance-based ab initio folding method predicts many more correct folds for the CASP12 targets than the best CASP12 group that integrated contacts, fragments, server predictions and conformation sampling into a complex folding simulation protocol. The rigorous experimental validation on 15 CASP13 targets shows that our distance-based folding servers successfully folded 2 targets (353 and 249 residues, respectively, with 148 and 120 effective sequence homologs) out of 3 hardest ones while the other servers failed on all. Our ab initio folding server also predicted the best, high-quality 3D model for a homology modeling target of 334 residues. The rigorous experimental validation in CAMEO shows that our ab initio folding server predicted a correct fold for a membrane protein of new fold while all the other CAMEO-participating servers failed. Finally, our distance-based algorithm runs very fast, folding the 41 CAMEO targets within 13 hours and the 37 CASP12 targets within 4 hours on a single Linux computer of 20 CPUs, excluding time for feature generation.

## Methods

## Datasets and Methods to Compare

We test our distance-based folding methods by 2 datasets: the set of 37 CASP12 hard targets[10, 11] and the set of 41 CAMEO hard targets[10]. Both datasets have been used to test our contact prediction algorithm before. CASP is a prestigious blind test of protein structure prediction organized by a human committee and CAMEO is an online blind test of structure prediction run by Schwede group[29]. Please see our papers [10, 11] for the detailed description of these two datasets. To make a fair comparison, we did not regenerate the features for the test proteins. Instead we used the same input features for them as described in[10, 11]. We trained our deep learning models using a set of ~10,000 non-redundant proteins deposited to PDB before May 2016 when CASP12 started. The CAMEO hard targets were collected in September and October 2016. This guarantees that the test targets have no similar proteins in the training set. We used 25% sequence identity as the cutoff to determine if two proteins are redundant or not. While training our deep learning model, we selected 600 proteins to form a validation set and the remaining ~9400 to form a training set. For the CASP12 dataset, we compared our method with CCMpred, MetaPSICOV and 4 top CASP12 groups. For the CAMEO dataset, we compared our method with CCMpred and MetaPSICOV. CCMpred is one of the best DCA (direct coupling analysis) methods, as reported in[10, 12]. MetaPSICOV is the best method for contact prediction in CASP11.

In addition, we have blindly tested our algorithm in the well-known blind test CASP13 in Summer 2018. As of November 1, 2018, 15 of ~90 CASP13 test targets have publicly available native structures, so we evaluate our performance in CASP13 using the 15 targets. For CASP13 targets, we compare our servers with all the other CASP13-participating servers.

## MSA (multiple sequence alignment) and protein feature generation

For the CASP12 and CAMEO datasets, to ensure fair comparison with the results in[10, 11] and the CASP12 groups, we used the same MSAs and protein features as described in[10, 11] for the test proteins and the same MSAs and protein features as described in[10] for the training proteins. That is, for each test protein we generated four MSAs by running HHblits with 3 iterations and E-value set to 0.001 and 1, respectively, to search the uniprot20 library released in November 2015 and February 2016, respectively. Since the sequence databases were created before CASP12 started in May 2016, the comparison with the CASP12 groups is fair. From each individual MSA, we derived both sequential and pairwise features. Sequential features include sequence profile and secondary structure and solvent accessibility predicted by RaptorX-Property[30]. Pairwise features include mutual information (MI), pairwise contact potential and direct information (DI) generated by CCMpred [6]. In summary, one test protein has 4 sets of input features and accordingly 4 predicted distance matrices, which are then averaged to obtain the final prediction.

For CASP13 targets, we generated their MSAs (and other sequence features) using the UniClust30 library[31] created in October 2017 and the UniRef sequence database[32] created early in 2018.

## Predict inter-atom distance by deep 1D and 2D deep residual networks (ResNet)

We use a very similar deep learning (DL) method as described in[10] to predict the Euclidean distance distribution of two atoms (of different residues) in a protein to be folded. The DL model in[10] is designed for contact prediction (i.e., binary classification) and formed by one 1D deep ResNet, one 2D deep ResNet and one logistic regression (see Fig. S1 in Appendix for a picture). ResNet is one type of DCNN, but much more powerful than traditional DCNN. The 1D ResNet is used to capture sequential context of one residue (or sequence motifs) while the 2D network is used to capture pairwise context of a residue pair (or structure motifs). For distance prediction, we replace the traditional convolutional operation in 2D ResNet by a dilated convolutional operation[33], which is slightly better since it needs fewer model parameters to have the same receptive field.

We discretize inter-atom distance into 25 bins: <4Å, 4-4.5Å, 4.5-5Å, 5-5.5Å, …, 15-15.5Å, 15.5-16Å, and >16Å. That is, we use 25 labels for distance prediction, as opposed to 2 labels for contact prediction. The DL model for distance prediction is trained using the same procedure as described in[10]. By summing up the predicted probability values of the first 9 distance labels (corresponding to distance≤8Å), our distance prediction DL model can also be used for contact prediction. Such a distance-based model has slightly better contact prediction accuracy than the DL model directly trained for contact prediction.

In addition to predict $C_\beta$-$C_\beta$ distance distribution, we also train individual DL models to predict distance distribution for the following atom pairs: $C_\alpha$-$C_\alpha$, $C_\alpha$-$C_g$, $C_g$-$C_g$, and N-O. Here $C_g$ represents the first CG atom in an amino acid. In the case that CG does not exist, OG or SG is used. The predicted distance of these 5 atom pairs is used together to fold a protein, which on average is better than using the predicted $C_\beta$-$C_\beta$ distance alone. The predicted distance for $C_\alpha$-$C_g$ and $C_g$-$C_g$ is also helpful for side chain packing.

## *Predict secondary structure and torsion angles by 1D deep residual network*

We employed a 1D deep ResNet of 19 convolutional layers to predict 3-state secondary structure and backbone torsion angles $\phi$ and $\psi$ for each residue. The 1D ResNet has the same architecture as what we used for contact/distance prediction[10] except the number of layers. Two types of input features are used: position specific scoring matrix (PSSM) generated by HHblits[34] and primary sequence represented as a 20×L binary matrix where L is sequence length. For secondary structure, logistic regression is used in the last layer to predict the probability of 3 secondary structure types. However, for torsion angles we do not use logistic regression, but directly predict the angle distribution defined as follows.

$$P(\phi,\psi \mid \bar{\phi},\bar{\psi},\sigma_1,\sigma_2,\rho) = \frac{1}{2\pi\sigma_1\sigma_2\sqrt{(1-\rho^2)}} \exp\{-\frac{1}{1-\rho^2}\left[\frac{1-\cos(\phi-\bar{\phi})}{\sigma_1^2} - \rho\frac{\sin(\phi-\bar{\phi})\sin(\psi-\bar{\psi})}{\sigma_1^2\sigma_2^2} + \frac{1-\cos(\psi-\bar{\psi})}{\sigma_2^2}\right]\} \quad (1)$$

In Eqn.(1), $\bar{\phi},\bar{\psi}$ are the mean, $\sigma_1,\sigma_2$ are the variance and $\rho$ is the correlation. That is, our deep ResNet outputs the mean and variance of the torsion angles associated with each residue. We use maximum-likelihood to train the network for secondary structure and angle prediction, i.e., maximizing the probability (defined by logistic regression or Eqn.(1) ) of the observed properties of our training proteins. Note that the predicted mean and variance for angles is residue-specific, i.e., each residue has a different pair of predicted mean and variance. Our method for angle prediction is different from many existing ones, which usually discretize angles into some bins and then formulate the problem as a classification problem. The same set of training proteins (created in 2016) for distance prediction is used to train our DL models for secondary structure and angle prediction.

## *Folding by predicted distance, secondary structure and torsion angles*

Given a protein to be folded, we first predict its inter-atom distance matrix, secondary structure and backbone torsion angles, then convert the predicted information into CNS restraints and finally build its 3D models by CNS[28], a software program for experimental protein structure determination.

Given a matrix of predicted distance probability distribution, for each atom type we first pick up 7L (L is sequence length) of residue pairs with the largest predicted probability of distance <15Å and assume them having distance<15Å. That is, we set their probability of having distance>16Å to be 0 and then estimate their distance. Given predicted distance probability distribution, we may estimate the mean distance and standard deviation (denoted as *m* and *s*, respectively) of one atom pair, and then use *m-s* and *m+s* as its distance lower and upper bounds. We used the same method as CONFOLD[35] to derive hydrogen bond restraints from predicted alpha helices. CONFOLD derived backbone torsion angles from predicted secondary structure, but we use the mean degree and variance predicted by our 1D deep ResNet as torsion angle restraints.

For each protein, we run CNS to generate 200 possible 3D decoys and then choose 5 with the least violation of distance restraints as the final models. CNS uses distance geometry to build 3D models from distance restraints and thus, can generate a 3D model within seconds. We generated multiple decoys for a protein since CNS solution is not globally optimal.

## The number of effective sequence homologs (Meff)

*Meff* measures the number of non-redundant (or effective) sequence homologs in an MSA. Here we use 70% sequence identity as cutoff to determine redundancy. Let a binary variable $S_{ij}$ denote if two protein sequences $i$ and $j$ are similar or not. $S_{ij}$ is equal to 1 if and only if the sequence identity between $i$ and $j$ is >70%. For a protein $i$, let $S_i$ denote the sum of $S_{i1}, S_{i2, ..., } S_{i,n}$ where $n$ is the number of proteins in the MSA. Then, *Meff* is calculated as the sum of $1/S_1, 1/S_2, …,1/S_n$. Meff measures the difficulty of contact prediction especially for the DCA methods. The smaller Meff a target has, the harder it is for contact prediction.

## Performance metrics

We mainly use TMscore[36] to evaluate the quality of a 3D model, which measures the similarity between a 3D model and its native structure (i.e., ground truth). It ranges from 0 to 1 and usually a 3D model is assumed to have a correct fold when it has TMscore≥0.5. Sometimes RMSD is also used to measure the deviation (in Å) of a 3D model from its native.

We measure the accuracy of predicted distance using 5 metrics: absolute error, relative error, precision, recall and *F1*. Since only predicted distance≤15Å is used as restraints to build 3D models, we evaluate only the atom pairs with predicted distance≤15Å. We define absolute error as the absolute difference between predicted distance and its native value, relative error as the absolute error normalized by the average of predicted distance and its native, precision as the percentage of atom pairs with predicted distance≤15Å that have native distance≤15Å and recall as the percentage of atom pairs with native distance≤15Å that have predicted distance≤15Å. *F1* is calculated by *2×precision×recall/(precision+recall)*.

# Results

## Results on the 37 CASP12 hard targets

Table 1 summarizes folding accuracy of various methods on the 37 CASP12 hard targets, including our distance-based ab initio folding method, three contact-based ab initio folding methods (i.e., our own method, CCMpred and MetaPSICOV) and 4 top CASP12 groups (Baker-server[37], Baker-human[38], Zhang-server[39] and Zhang-human[40]). See Supplementary File 1 for the detailed modeling accuracy. As shown in Table 1, the 3D models predicted by our distance-based ab initio folding method have average TMscore 0.466 and 0.476, respectively, when the top 1 and the best of top 5 models are evaluated, much better than the models built from contacts predicted by ourselves, CCMpred and MetaPSICOV. Our distance-based 3D models also have much higher quality than the 4 top CASP12 groups, which folded proteins by combining fragment assembly, predicted contacts, conformation sampling and energy minimization. The two human groups also extracted information from the predictions of all CASP12-participating servers. By contrast, our folding algorithm does not use any template fragments or any CASP12 server predictions.

Table 1. The average quality (TMscore) of the 3D models predicted by various methods for the 37 CASP12 hard targets. Top k (k=1, 5) means that for each target the best of top k 3D models is considered. Columns "Top 1" and "Top 5" list the average TMscore of the top 1 and the best of top 5 models for all the targets, respectively. Column "#correct folds" lists the number of targets with correctly predicted folds (TMscore>0.5) when all 5 models are considered. The results in rows 3-9 are copied from our previous paper [11]. See Supplementary File 1 for details.

| Group | Top 1 | Top 5 | #correct folds |
|---|---|---|---|
| **This work** | 0.466 | 0.476 | 21 |
| Our contact | 0.354 | 0.397 | 10 |
| CCMpred | 0.216 | 0.235 | 0 |
| MetaPSICOV | 0.262 | 0.289 | 2 |
| Baker-server | 0.326 | 0.370 | 9 |
| Zhang-server | 0.347 | 0.404 | 10 |
| Baker-human | 0.392 | 0.422 | 11 |
| Zhang-human | 0.375 | 0.420 | 11 |

When all 5 models are considered, our distance-based folding can predict correct folds (TMscore>0.5) for 21 of the 37 targets while our contact-based folding can do so for only 9 of them and the best CASP12 human group can do so for only 11 of them. See Figure 1(A) for the detailed comparison between our distance-based and contact-based models. With only predicted contacts and predicted secondary structure, MetaPSICOV generates correct fold for only 2 targets and CCMpred cannot generate any correct folds. That is, the contacts predicted by CCMpred and MetaPSICOV alone are not good enough for 3D modeling.

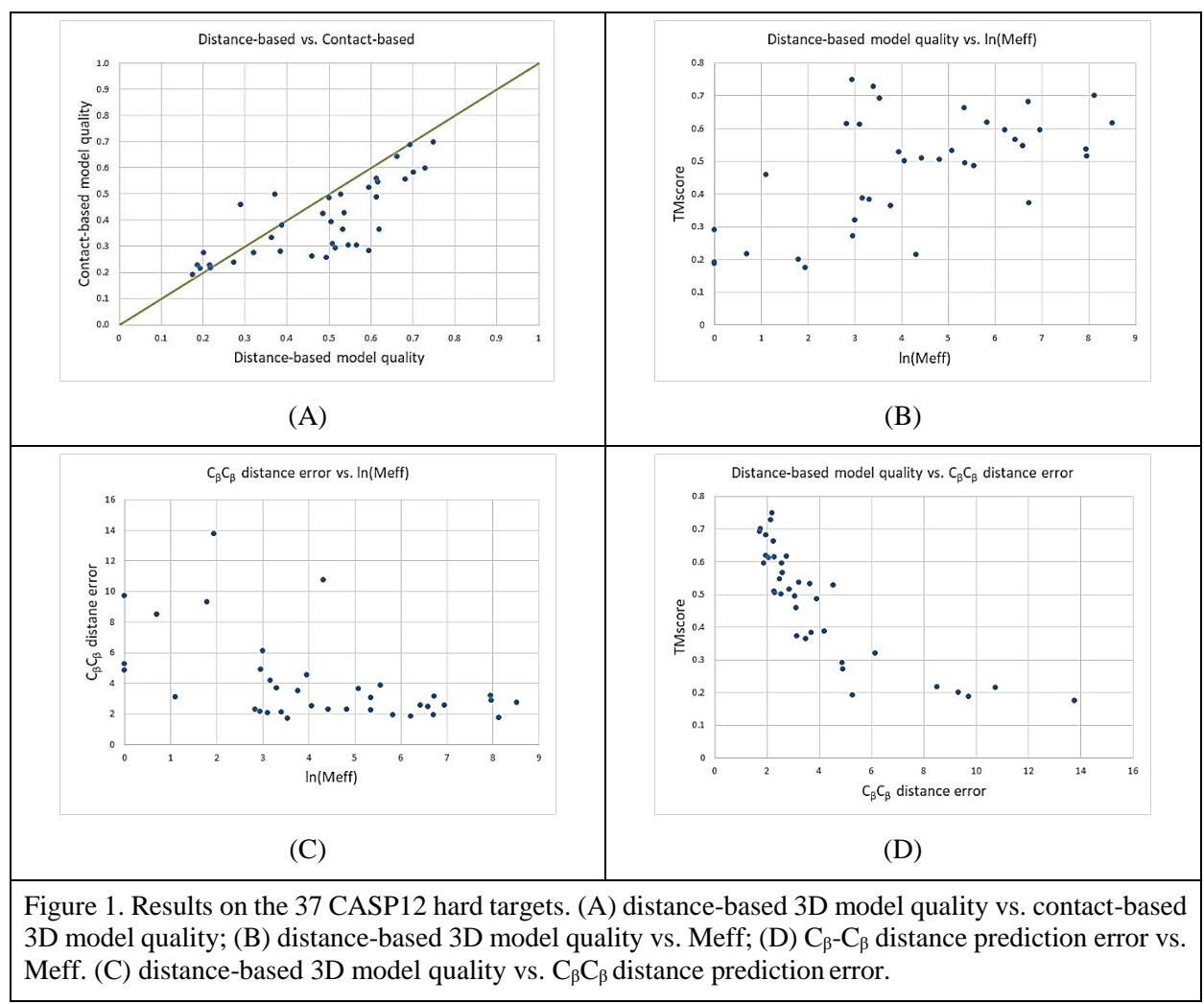

Figure 1. Results on the 37 CASP12 hard targets. (A) distance-based 3D model quality vs. contact-based 3D model quality; (B) distance-based 3D model quality vs. Meff; (D) $C_\beta$-$C_\beta$ distance prediction error vs. Meff. (C) distance-based 3D model quality vs. $C_\beta C_\beta$ distance prediction error.

Figure 1(B) shows that the quality of our distance-based 3D models is correlated to Meff, which measures the number of sequence homologs of a target. When Meff>55 or ln(Meff)>4, there is a good chance that our predicted 3D models have a correct fold. Our distance-based ab initio folding can fold 8 out of 21 targets

with Meff≤100: T0862-D1, T0863-D1, T0869-D1, T0870-D1, T0894-D1, T0898-D1, T0904-D1 and T0915-D1. Meanwhile, 5 of them have 3D models with TMscore>0.6. In contrast, Zhang-Server, Zhang-Human, Baker-Server and Baker-Human predicted models with TMscore>0.6 for only 2, 1, 0 and 0 targets with Meff≤100, respectively.

**Evaluation of distance prediction.** Here we only consider the pairs of atoms with sequence separation at least 12 residues and predicted distance≤15Å. Table 2 summarizes the quality of predicted distance on the 37 hard CASP12 targets. The quality is first calculated on each target and then averaged across all the test targets. See Supplementary File 1 for the details. Figure 1(C) shows that there is certain correlation between distance prediction error and Meff. Figure 1(D) shows a strong correlation between 3D modeling quality and distance prediction error, which means that as long as distance prediction is accurate, CNS is able to build good 3D models. When the distance error is 8Å, the predicted 3D models are equivalent to a random model in terms of quality.

Table 2. Quality of predicted distance on the 37 hard CASP12 targets.

| Atom pair | Absolute error | Relative error | Precision | Recall | F1 |
|---|---|---|---|---|---|
| $C_\beta C_\beta$ | 4.045Å | 0.269 | 0.6541 | 0.5106 | 0.5677 |
| $C_g C_g$ | 4.498Å | 0.296 | 0.6123 | 0.5153 | 0.5529 |
| $C_\alpha C_g$ | 4.421Å | 0.285 | 0.6323 | 0.4955 | 0.5485 |
| $C_\alpha C_\alpha$ | 3.938Å | 0.257 | 0.6499 | 0.5229 | 0.5733 |
| NO | 4.035Å | 0.262 | 0.6412 | 0.4996 | 0.5563 |

## Results on the 41 CAMEO hard targets

On average, this dataset is easier than the CASP12 set. Again, we build 3D models using CNS from the predicted distance and angle restraints. Our folding algorithm does not use any template fragments or predictions produced by any other servers. See Supplementary File 1 for our detailed modeling accuracy. In summary, the result on this dataset is consistent with that on the CASP12 test set. When the first models and the best of top 5 models are evaluated, our distance-based ab initio folding algorithm has average TMscore 0.551 and 0.577, respectively, about 10% better than our contact-based ab initio folding algorithm, which has average TMscore 0.504 and 0.524, respectively. Figure 2(A) shows that for most targets, our distance-based models have better quality than our contact-based models. In total, our distance- and contact-based methods predicted correct folds for 30 and 23 targets, respectively. Figure 2(B) shows that our distance-based model quality is correlated to Meff (i.e., the number of sequence homologs available for a target), but our method predicted a correct fold for 1 target with Meff=1. Similar to the observation on the CASP12 dataset, when Meff>55 or ln(Meff)>4, there is a good chance that our predicted 3D models have a correct fold. Figure 2(D) shows that there is strong correlation between 3D modeling accuracy and distance prediction error. Finally, when all top 5 models are considered, the 3D models generated by CCMpred contacts and MetaPSICOV contacts have average TMscore 0.316 and 0.392, respectively, much worse than our distance-based method.

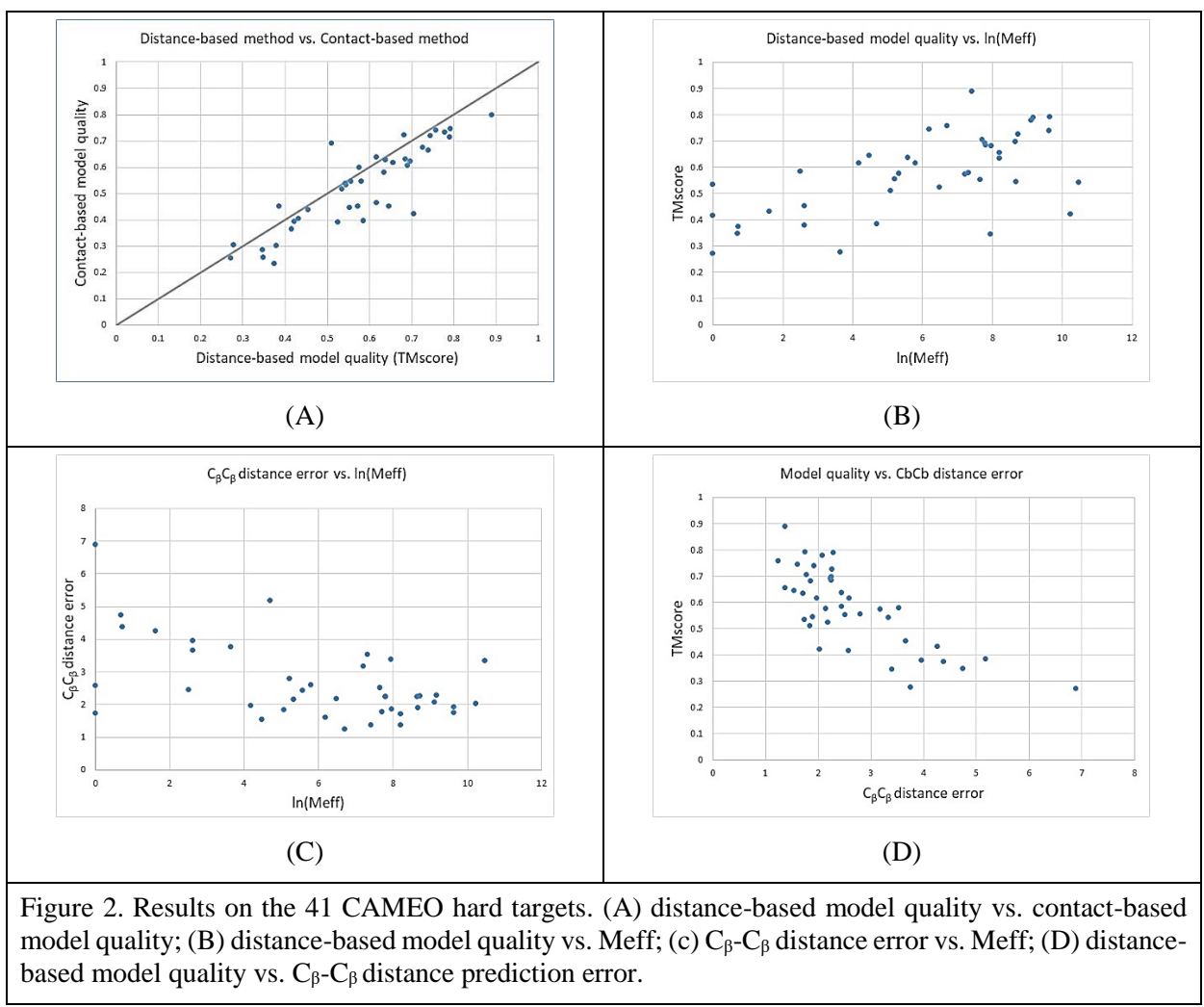

Figure 2. Results on the 41 CAMEO hard targets. (A) distance-based model quality vs. contact-based model quality; (B) distance-based model quality vs. Meff; (c) $C_\beta$-$C_\beta$ distance error vs. Meff; (D) distance-based model quality vs. $C_\beta$-$C_\beta$ distance prediction error.

**Evaluation of distance prediction.** Here we only consider the pairs of atoms with sequence separation at least 12 residues and predicted distance≤15Å. Table 3 summarizes the average quality of predicted distance on the 41 CAMEO hard targets. See Supplementary File 1 for the quality on each target. Figure 2(C) shows that the distance prediction error is inversely proportional to the logarithm of Meff. When Meff>55 or ln(Meff)>4, there is a very good chance that the average $C_\beta$-$C_\beta$ distance prediction error is less than 4Å. This is consistent with what is shown in Figure 1(C).

**Table 4. Evaluation of distance prediction on the 41 CAMEO hard targets.**

| Atom pair | Absolute error | Relative error | Precision | Recall | F1 |
|---|---|---|---|---|---|
| $C_\beta C_\beta$ | 2.725Å | 0.207 | 0.6941 | 0.6133 | 0.6158 |
| $C_g C_g$ | 3.212Å | 0.237 | 0.6580 | 0.6074 | 0.5921 |
| $C_\alpha C_g$ | 3.058Å | 0.223 | 0.6687 | 0.5653 | 0.5914 |
| $C_\alpha C_\alpha$ | 2.726Å | 0.197 | 0.7032 | 0.5943 | 0.6230 |
| NO | 2.814Å | 0.196 | 0.6987 | 0.5770 | 0.6139 |

## Rigorous experimental validation in CASP13

We have validated our distance-based folding algorithm in the rigorous blind test CASP13 in Summer 2018. While doing CASP13, the native structures of the test proteins have not been solved yet. As of November 1, 2018, 15 of ~90 CASP13 targets have publicly available native structures. We registered three servers: RaptorX-Contact for both contact prediction and distance-based ab initio folding, RaptorX-TBM for pure protein threading[27] and RaptorX-DeepModeller for distance-based folding of a target with only remote templates. RaptorX-Contact is designed for very hard targets without templates, RaptorX-TBM for easy targets with good templates and RaptorX-DeepModeller for targets with weakly similar templates. RaptorX-TBM used Rosetta to build 3D models from alignments generated by our own threading method. When a test target has a good template (i.e., sequence identity≥30%), RaptorX-DeepModeller used the RaptorX-TBM models. Otherwise RaptorX-DeepModeller used our distance-based folding method to build a 3D model, but with a different ResNet model. In addition to those input features used by the ResNet model for RaptorX-Contact, the ResNet model for RaptorX-DeepModeller has one more input feature, namely, an initial distance matrix extracted from the template according to the alignment. Supposing two residues of the query protein i and j are aligned to residues k and l of the template, we assign the distance between k and l as the initial distance of i and j. When one query residue is not aligned, the corresponding row and column in the initial distance matrix is empty and we will use a binary flag to indicate this. When the template is not good, the 3D models produced by RaptorX-DeepModeller may be very different from the RaptorX-TBM models, although they use the same alignment.

The official CASP13 results will not be released until early in December 2108. Among the 15 targets with publicly available native structures, 6 are hard: T0950, T0953s1, T0953s2, T0958, T0960 and T0963. Their Meff values are 148, 97, 120, 18, 237, and 340, respectively. Meanwhile, T0958 has a weakly similar template in PDB. T0960 and T0963 are multi-domain proteins; one of their domains has a reasonable template and the other domains may not. That is, among these 6 hard targets, T0950, T0953s1 and T0953s2 may be the most challenging. Since we do not have the official domain definitions, here we use the whole chain of a target as an evaluation unit.

RaptorX-Contact has the best contact prediction accuracy on the 6 hard targets. When the top L/2 long-range contact predictions are evaluated, the top 5 servers (accuracy) are: RaptorX-Contact (64.0%), 189 (55.6%), ResTriplet (54.3%), TripletRes (52.8%), and 491 (52.3%). When the top L/2 medium-range contact predictions are evaluated, the top 5 servers (accuracy) are: RaptorX-Contact (55.0%), TripletRes (51.2%), Yang-Server (50.0%), ResTriplet (49.6%) and 491(47.2%). To the best of our knowledge, in addition to RaptorX-Contact, the other top servers also used ResNet in CASP13.

Our servers did very well in predicting 3D models for hard targets. Here we evaluate the first models submitted by all CASP13-participating servers, which used a variety of prediction methods and many of them have used predicted contacts. Among the 6 hard targets, all the servers failed to predict correct folds for 3 hard targets: T0953s1, T0960 and T0963. The best TMscore for T0953s1 is less than 0.4 and the best for T0960 and T0963 is less than 0.3. The predicted models for T0960 and T0963 have low TMscore mainly because they are multi-domain proteins and it is more reasonable to evaluate them by domains. Below are the results of the other 3 targets:

1) For T0950 (an α protein of 353 residues and Meff=148), the top 5 servers (TMscore) are: RaptorX-DeepModeller (0.5632), Baker (0.4642), QUARK (0.4444), Zhang-Server (0.4405) and RaptorX-TBM (0.4366). RaptorX-Contact produced a model with TMscore 0.3790. When TMscore=0.5 is used as a cutoff to judge if a 3D model has a correct fold, only RaptorX-DeepModeller generated a correct fold for T0950 while all the other servers failed. RaptorX-DeepModeller achieved this by mixing information used by RaptorX-TBM and RaptorX-Contact, although the latter two failed to produce a correct fold.
2) For T0953s2 (a β protein of 249 residues and Meff=120), the top 5 servers (TMscore) are: RaptorX-Contact (0.5677), RaptorX-DeepModeller (0.4771), AWSEM-Suite (0.4631), QUARK (0.4588) and

Zhang-Server (0.4453). That is, only RaptorX-Contact generated a correct fold for T0953s2 while all the others failed.

3) T0958 (an α/β protein of 96 residues and Meff=18) has a weakly similar template. The top 5 servers (TMscore) are: AWSEM-Suite (0.6944), RaptorX-DeepModeller (0.6373), Seok-Server (0.6196), RaptorX-TBM (0.6061), and Zhang-CEthreader (0.6020). Although not among top 5, the RaptorX-Contact model has a correct fold (TMscore=0.5294), which is significant given that T0958 has only 18 effective sequence homologs. In addition, AWSEM-Suite [41] used our predicted contacts for CASP13.

In summary, our group predicted correct folds for 2 large, hardest targets (T0950 and T0953s2) with <150 effective sequence homologs while the others got none correct. RaptorX-Contact is a pure ab initio folding server, so it may not compete with the top servers on easy targets. Nevertheless, RaptorX-Contact predicted the best 3D model (TMscore=0.8442) for an easy target T0965 (an α/β protein of 334 residues) with very good templates. This is interesting since an easy target is usually predicted by homology modeling plus refinement, but this example shows that ab initio folding may work better. By the way, when all the 15 targets are considered, RaptorX-Contact still has the best contact accuracy and RaptorX-DeepModeller has the best overall TMscore.

## Rigorous experimental validation in CAMEO

Starting from middle September 2018, we have tested our distance-based ab initio folding web server (http://raptorx.uchicago.edu/AbInitioFolding/) in the online benchmark CAMEO (http://www.cameo3d.org/) operated by Schwede group. CAMEO is a blind test and the native structure of a test protein is usually released one week after the test. Currently, CAMEO is testing ~40 web servers including some popular ones such as Robetta, Phyre, RaptorX, Swiss-Model, IntFold and HHpred. Here we only show the result of one recent target (CAMEO-3D ID: 2018-11-03_00000053_1) which is a membrane protein with a new fold since CASP13 does not seem to have this kind of target. This target has a PDB ID 6bhp and chain name C, 200 residues and 229 effective sequence homologs. As shown in Fig. 3, the best 3D model generated by our ab initio folding server (ID: server 60) has TMscore=0.68 and RMSD=5.65Å, much better than those generated by the other servers, which failed to predict a correct fold. Further, our response time is much shorter than that of Baker's Robetta. Meanwhile, this job was queued in our server for about 4.5 hours and the actual time spent in folding this target is about 1.5 hours.

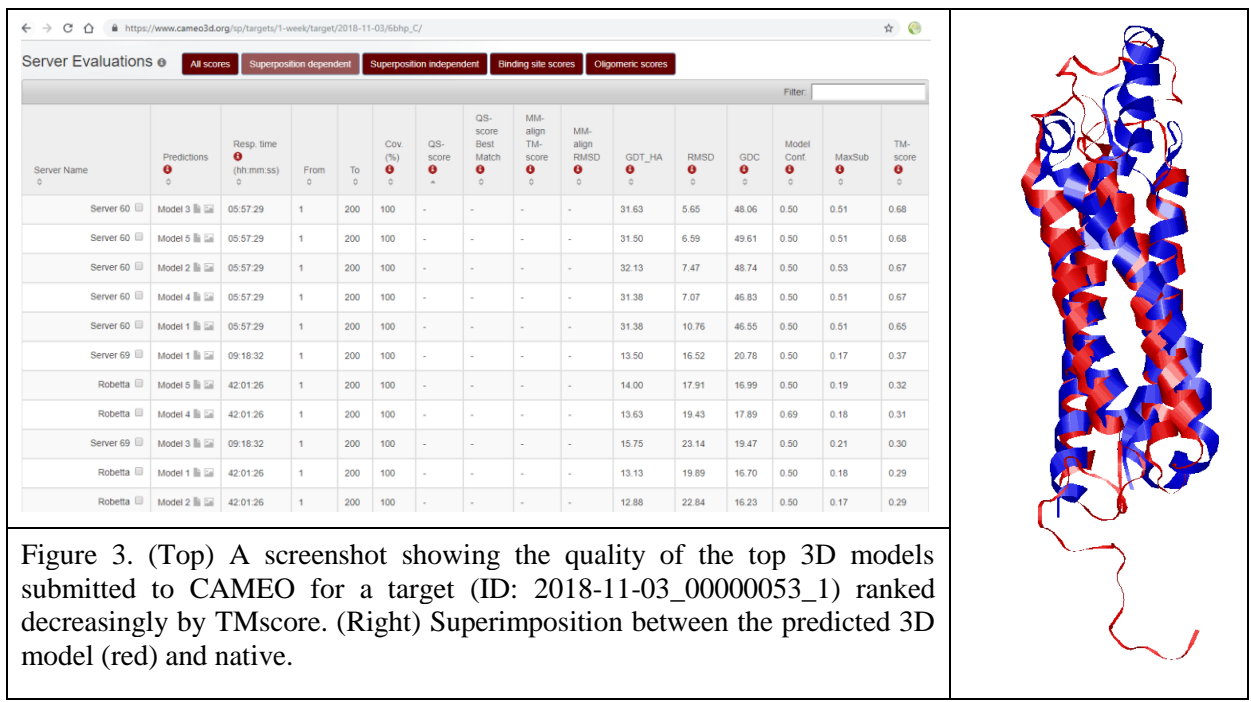

Figure 3. (Top) A screenshot showing the quality of the top 3D models submitted to CAMEO for a target (ID: 2018-11-03_00000053_1) ranked decreasingly by TMscore. (Right) Superimposition between the predicted 3D model (red) and native.

## Running time

Our distance-based ab initio folding algorithm runs very fast since it does not involve folding simulation. The whole folding pipeline consists of three main steps: 1) generating multiple sequence alignments and input features; 2) predicting angles and distance; and 3) folding by feeding distance and angle restraints to CNS. It takes minutes to finish the first step for most targets, seconds to finish the second step on a GPU card, and 10 minutes to a few hours to finish the third step on a Linux computer of 20 CPUs. By running the third step in parallel on a Linux computer of 20 CPUs, it took in total ~13 hours to fold the 41 CAMEO targets and ~4 hours to fold the 37 CASP12 targets.

## Conclusion and Discussion

We have shown that we can predict inter-atom distance matrix very well and that without folding simulation, distance-based protein folding can yield much better 3D models than contact-based folding. Although our algorithm does not use fragment assembly, complex energy function and time-consuming folding simulation, it can predict correct folds for many more hard targets than the best CASP12 groups. Rigorous experimental validation in CASP13 and CAMEO confirmed that our distance-based folding algorithm works for some very hard targets while the other servers failed. Since it does not need folding simulation at all, our algorithm runs very fast, taking from 10 minutes to a few hours to generate 200 decoys on a Linux computer of 20 CPUs. That is, we may do ab initio folding on a laptop equipped with a GPU card.

In this study we do not evaluate in detail the accuracy of our 1D deep residual neural network models for secondary structure and torsion angle prediction because 1) the prediction accuracy is similar to what we have reported before[42, 43] although the methods are different and 2) secondary structure and torsion angles are much less important than distance for protein folding. Without using predicted torsion angles, on average the 3D model quality decreases by ~0.008 in terms of TMscore. Nevertheless, the predicted torsion angles may help reduce the number of mirror images generated by CNS.

In this study we only reported the folding results when all the 5 types of atom pairs are used. In fact, using only $C_\beta$-$C_\beta$, our distance-based ab initio folding algorithm can generate slightly worse 3D models than using all 5 types of atom pairs because their distance is highly correlated. Among the 5 types of atom pairs, $C_\beta$-$C_\beta$ is the most useful when only the $C\alpha$ conformation is evaluated. Nevertheless, using all 5 types of atom pairs can help reduce noise a little bit and may improve side chain packing.

In this study we predicted inter-residue distance by discretizing it into 25 bins. We have also experimented with discretizing the distance into 12 bins (i.e., bin width=1Å) and 52 bins (i.e., bin width=0.25Å). On average, there is no big difference between using 25 bins and 52 bins, both of which are better than using 12 bins. Instead of using a discrete representation of distance, we may predict real-valued distance matrix by assuming that distance has a log-normal distribution[44] and then revise our deep learning model to predict the mean and variance of the distribution. CNS can easily take the predicted mean and variance as distance restraints to build 3D models. Currently we use only distance≤15Å for 3D model construction. We will study if 3D modeling accuracy can be further improved by using the whole real-valued distance matrix, especially for the determination of domain orientation of a multi-domain protein.

Recently an end-to-end deep learning method was proposed by Dr. AlQuraishi that can directly predict 3D coordinates of a protein structure from its sequence profile[45]. Our ResNet method is also end-to-end, predicting inter-atom distance matrix from sequence profile, mutual information and direct information. Although our ResNet model does not directly yield 3D coordinates, it is not difficult to construct protein 3D models from predicted distance matrix using CNS. In the rigorous CASP13 test, we have not observed that Dr. AlQuraishi's group (HMSCasper-Refine) has succeeded on any of the three hardest targets (T0950, T0953s1, T0953s2).

Finally, it will be interesting to see if 3D modeling accuracy can be further improved if our predicted distance is integrated into a fragment-based folding simulation pipeline such as Rosetta, although this may significantly increase running time.

## Acknowledgement

This work is supported by National Institutes of Health grant R01GM089753 to JX and National Science Foundation grant DBI-1564955 to JX. The authors are also grateful to the support of Nvidia Inc. The funders had no role in study design, data collection and analysis, decision to publish, or preparation of the manuscript.

The author is grateful to postdoc Dr. Sheng Wang, who helped generate multiple sequence alignment and input features for the CASP13 targets.

# Appendix

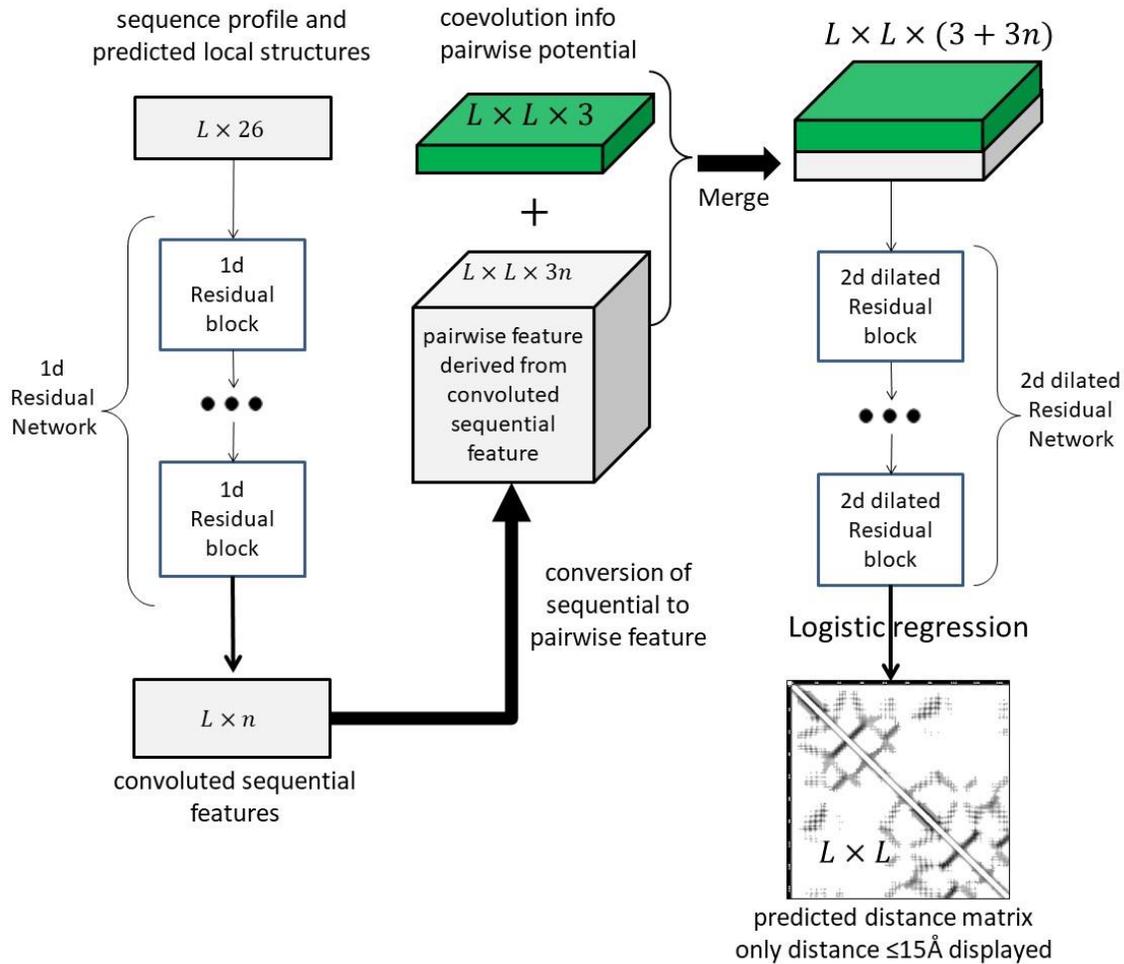

Figure S1. The overall deep network architecture for the prediction of protein distance matrix. The left column is a 1D deep residual neural network that transforms sequential features (e.g., sequence profile and predicted secondary structure). The right column is a 2D deep dilated residual neural network that transforms pairwise features. The middle column converts the convoluted sequential features to pairwise features and combine them with the original pairwise features. The picture is adapted from Figure 1 in the paper at https://journals.plos.org/ploscompbiol/article?id=10.1371/journal.pcbi.1005324 .